\documentclass[useAMS,usenatbib]{mn2e}
\usepackage{url,times,graphicx,amsmath,amsfonts,amssymb,aas_macros,color,epsfig,epstopdf}

\newcommand{\mockalph}[1]{}
\newcommand{\hMpc}{{\ifmmode{h^{-1}{\rm Mpc}}\else{$h^{-1}$Mpc}\fi}}
\newcommand{\hkpc}{{\ifmmode{h^{-1}{\rm kpc}}\else{$h^{-1}$kpc}\fi}}
\newcommand{\hMsun}{{\ifmmode{h^{-1}{\rm {M_{\odot}}}}\else{$h^{-1}{\rm{M_{\odot}}}$}\fi}}
\newcommand{\ltsima}{$\; \buildrel < \over \sim \;$}
\newcommand{\gtsima}{$\; \buildrel > \over \sim \;$}
\newcommand{\lsim}{\lower.5ex\hbox{\ltsima}}
\newcommand{\gsim}{\lower.5ex\hbox{\gtsima}}
\def\LCDM{$\Lambda$CDM}

\def\lesssim{\mathrel{\hbox{\rlap{\hbox{\lower4pt\hbox{$\sim$}}}\hbox{$<$}}}}
\def\gtrsim{\mathrel{\hbox{\rlap{\hbox{\lower4pt\hbox{$\sim$}}}\hbox{$>$}}}}
\def\lsim{\mathrel{\rlap{\lower3pt\hbox{$\sim$}}
    \raise1pt\hbox{$<$}}}                
\def\gsim{\mathrel{\rlap{\lower3pt\hbox{$\sim$}}
    \raise1pt\hbox{$>$}}}                

\newcommand{\mstar}{M$_{\rm star}$}

\newcommand{\msun}{M$_\odot$}

\newcommand{\mhalo}{M$_{\rm halo}$}

\newcommand{\rhalf}{r$_{\rm half}$}

\newcommand{\Sec}[1]{Section~\ref{#1}}
\newcommand{\Eq}[1]{Eq.~(\ref{#1})}

\newcommand{\beq}{\begin{equation}}
\newcommand{\eeq}{\end{equation}}
\def\beqa{\begin{eqnarray}}
\def\eeqa{\end{eqnarray}}
\def\hMpc{$h^{-1}\,{\rm Mpc}$}
\def\hkpc{$h^{-1}\,{\rm kpc}$}
\def\LCDM{\ensuremath{\Lambda}CDM}

\def\head{
 \vbox to 0pt{\vss
                   \hbox to 0pt{\hskip 440pt\rm LA-UR-10-07069\hss}
                  \vskip 25pt}}

\title[Too-big-to-fail and core/cusp problem]
{Expanded haloes, abundance matching and too-big-to-fail in the Local Group}
\author[Brook \& Di Cintio]
       {Chris B. Brook$^{1}$\thanks{E-mail: cbabrook@gmail.com} \& Arianna Di Cintio$^{1,2}$\thanks{E-mail: arianna.dicintio@uam.es}\\
$^{1}$Departamento de F\'isica Te\'orica, M\'odulo C-15, Facultad de Ciencias, Universidad Aut\'onoma de Madrid, 28049 Cantoblanco, Madrid, Spain\\
$^{2}$Dark Cosmology Centre, Niels Bohr Institute, University of Copenhagen, Juliane Maries Vej 30, DK-2100 Copenhagen, Denmark\\}

\begin{document}

\date{Accepted XXXX . Received XXXX; in original form XXXX}

\pagerange{\pageref{firstpage}--\pageref{lastpage}} \pubyear{2010}

\maketitle

\label{firstpage}


\begin{abstract}
Observed kinematical data of 40 Local Group (LG) members are used to derive the dark matter halo mass of such galaxies.
Haloes are selected from the theoretically expected LG mass function and two different density profiles are assumed, a standard universal cuspy model and a mass dependent profile which accounts for the effects of baryons in modifying the dark matter distribution within galaxies. 
The resulting relations between  stellar and halo mass  are compared with expectations from abundance matching. 

Using a universal cuspy profile, the ensemble of LG galaxies is fit in relatively low mass haloes, leaving  ``dark" many massive haloes of \mhalo$\gtrsim$10$^{10}$\msun: this reflects the ``too big to fail" problem and results in a \mstar-\mhalo\ relation that differs from abundance matching predictions. Moreover, the star formation efficiency of isolated LG galaxies increases with decreasing halo mass when adopting a cuspy model.
By contrast, using  the mass dependent density profile, dwarf galaxies with \mstar$\gtrsim$10$^{6}$\msun are assigned to more massive haloes, which have a central cored distribution of dark matter: the ``too big to fail" problem is alleviated, the resultant  \mstar-\mhalo\ relation follows abundance matching predictions down to the completeness limit of current surveys, and the star formation efficiency of isolated members decreases with decreasing halo mass, in agreement with theoretical expectations.

Finally, the cusp/core space of LG galaxies is presented, providing a framework to understand the non-universality of their density profiles.

\end{abstract}
\noindent
\begin{keywords}
 galaxies: formation - haloes - Local Group - dwarf
 \end{keywords}
   
\section{Introduction} \label{sec:introduction}

The $\Lambda$ Cold Dark Matter (\LCDM) model has been  successful in explaining a multitude of observations at cosmological scales, such as anisotropies of Cosmic Microwave Background radiation (CMB) \cite[e.g.][]{Jarosik11} and galaxy clustering on large scales \citep[e.g.][]{Cole05}. However,  the \LCDM\ model  has  problems on galactic scales, such as the ``missing satellite problem", the ``too big to fail" problem and the ``cusp-core" discrepancy.
At these small scales,  tests of the \LCDM\ model are complicated by astrophysical processes involved in galaxy formation. 

The  ``missing satellite problem", which is the discrepancy between the number of observed satellites and  the number of predicted dark matter sub-haloes \citep{Klypin99,Moore99}, can be resolved if the lowest mass dark matter haloes are inefficient at forming stars due to the early reionization of the intergalactic medium  \citep{Bullock00,Somerville02,Benson02}.

Yet there remains a mismatch between the predicted and observed kinematics of galaxies in the mass range where haloes are  too massive to have star formation suppressed by reionization processes, i.e. they are  ``too big to fail" \citep{Boylan11}. A lack of observed galaxies with the kinematics expected for such haloes has been reported in satellites of the Milky Way (MW) and Andromeda (M31) \citep{Boylan12,collins14,tollerud14},  in the Local Group \citep{ferrero12,gk14,kirby14} and in the velocity function within the local volume, as measured by HI line widths \citep{klypin14,papastergis14}. High resolution rotation curves of dwarf  galaxies from the THINGS survey \citep{oh11} and a sample of low mass (10$^7<\rm M_{\rm star}/M_{\odot}<10^9$) star-forming galaxies at intermediate redshift \citep{miller14} show a similar disagreement with theoretical expectations, since their observed kinematics indicate that their haloes are less massive than abundance matching \citep{guo10,moster13} would suggest.

These discrepancies between theory and observation  arise under the assumption that  galaxies reside in dark matter haloes whose properties are derived from collisionless N-body cosmological simulations, i.e. their density profiles are steep, or ``cuspy", toward the centre \citep{navarro97}. 
Such NFW profile implies that  rotation curves of galaxies should increase  rapidly with radius.
However, there is ample evidence that observed rotation curves rise slowly, and that dwarf galaxies have flat, or ``cored", inner density profiles  \citep[e.g.,][]{moore94,Salucci00,deBlok01,Simon05,Kuzio08,oh11b}. This is the well known ``cusp-core" discrepancy. 

Within a \LCDM\ scenario, the ``too big to fail" problem is  likely a re-casting of the ``cusp-core" discrepancy, with the mismatch between observed and theoretical velocities of  massive dwarfs and satellites reflecting the existence of cores in such galaxies. 
Therefore, a possible solution to both the ``cusp-core" discrepancy  {\it and} the ``too big to fail" problem is the formation of cores through the effects of  baryonic physics \citep{governato12,madau14}, such as the non-adiabatic impact of gas outflows on dark matter haloes \citep{navarro96,read05,mashchenko08,pontzen12,Ogiya14}

There is significant observational evidence \citep{weiner09,martin12} that energy feedback from  star-formation activity  drives gas out of galaxies. Processes such as  radiation energy from massive stars,  stellar winds and supernova explosions  have been shown to play a central role in  galaxy formation \citep{binney01,brook11,stinson13,hopkins14}.

Both simple analytic arguments (Brook \& Di Cintio 2015 in prep) and detailed cosmological simulations \citep{DiCintio2014a} show that the degree of halo expansion due to outflows  is dependent on the ratio of stellar to halo mass, \mstar/\mhalo. Low mass galaxies are dark matter dominated: those with \mstar$\lsim$3-5$\times$10$^6$M$_\odot$ do not produce enough energy to flatten the halo's inner density profile, which remains steep \citep{penarrubia12,governato12,DiCintio2014a}.  As stellar mass increases relative to the dark matter mass,  the inner density profile becomes increasingly flat \citep{governato12,DiCintio2014a}. The flattening is greatest when   \mstar$\sim$3$\times$10$^8$\msun, after which the deeper potential well is able to  oppose the halo expansion \citep{DiCintio2014a,DiCintio2014b}, resulting in a profile which becomes steeper for more massive galaxies, and effectively returns to the NFW value at about the Milky Way mass.

While some scatter in the relation between inner slope and \mstar/\mhalo\ is certainly expected, due to the different evolution of galaxies, different star formation, merger histories and environments, at a first order the amount of stars per halo mass at redshift zero gives a good approximation of the energy from supernovae feedback, available to flatten the dark matter profiles, versus the total gravitational potential energy of the NFW halo.

Using the results from hydrodynamical galaxy formation simulations that match a wide range of galaxy scaling relations \citep{brook12,stinson13} and their evolution \citep{kannan14,obreja14}, we have parametrized, in terms of the ratio \mstar/\mhalo, a density profile that accounts for the effects of outflows in flattening the central dark matter distribution of haloes \citep{DiCintio2014b}.
This mass dependent profile  is fully described in terms of the ratio \mstar/\mhalo\ and we will refer to it, throughout the paper, as the DC14 profile.

We aim to test whether such a mass dependent density profile is able to account for the kinematics of Local Group (LG) galaxies and whether this helps reconciling the ``too big to fail" problem with abundance matching predictions.
We use observed stellar velocity dispersions of LG galaxies from \citet{kirby14}, which provide an  estimate of the mass M$(r_{1/2})$ (or equivalently of the circular velocity V$(r_{1/2})$) enclosed within the galaxy's half-light radius $r_{1/2}$, and fit such data with theoretical rotation curves, in order to find the best fit halo mass for each galaxy. To model the distribution of dark matter within haloes, we use both the mass dependent DC14 density profile and the common NFW one, and compare the results arising from these two models against abundance matching relations in the LG \citep{brook14,elvis}.

The study of galaxies and satellites in the LG has two advantages over studying satellite populations in  the Milky Way or M31: firstly, there is an added independent mass constraint coming from the timing argument \citep{Kahn59,LiWhite08}; secondly, the mass function approaches a power law  for all but the 10 most massive galaxies \citep{brook14},  which alleviates the large cosmic variance in satellite mass functions of individual haloes of the mass of M31 and the Milky Way \citep[e.g.][]{purcell12}. Nevertheless, cosmic variance remains an issue in matching haloes to luminous galaxies within the LG, primarily in the normalisation of the adopted power law. The LG mass is dominated by the combined mass of the M31 and the MW, which are not constrained to fit the extended power law that we adopt for the halo mass function. We do explore the effects of a low normalisation of the power law in section~\ref{NORM_LG} but it is beyond the scope of this paper to test whether an even lower normalisation is possible, a task which requires a significant number of LG analogue volumes to be simulated.   This issue, and the key question of the number of ``too big to fail" haloes,  was examined in detail in \cite{elvis}, to which we refer the reader.

The paper proceeds as follows: we present the observational data, the halo mass function, the abundance matching relation and  the mass dependent density profile in section~\ref{methods}.  We then use the observed kinematics to find the best fit dark matter halo for each galaxy in section~\ref{match}. We show the resultant relations between \mstar\ and \mhalo\ for the two different density profiles in section~\ref{MsMh}, comparing them to the expectations from  abundance matching and number of ``too big to fail" haloes in section~\ref{NORM_LG}. We discuss the implications of our findings in terms of star formation efficiency in the LG in section~\ref{sfe}. We present the cusp/core space of LG galaxies in section~\ref{color}, providing a framework for comparisons with observations.
In section~\ref{conclusion} we discuss how the NFW profile cannot satisfactorily explain the kinematics of the full population of LG galaxies when compared with abundance matching predictions, enhancing the ``too big to fail" problem, and how these issues are instead solved once the mass dependent DC14 profile is considered. We finally discuss possible caveats and implications of our findings.

\section{Data and Methods}\label{methods}
\subsection{Strategy}

Kinematic information of LG  members, both satellites and isolated, are used to find the mass of the dark matter halo that best fits each observed galaxy. 
The mass of each halo, \mhalo, is defined as the mass of a sphere of radius $\rm R_{\rm vir}$ containing $\Delta_{\rm vir}$ times the critical matter density of the Universe $\rho_{crit}=3H^2/8\pi G$ at z=0, where $\Delta_{\rm vir}=18\pi^2 + 82x - 39x^2$ \citep{Bryan98} and $x=\Omega_{m}-1$ at z=0 depends on the selected cosmology.

Two different density profiles are used for dark matter haloes, along with a cosmologically motivated power law halo mass function to derive the distribution of available halo masses to which LG galaxies are assigned.

\subsection{Halo mass function in the Local Group}\label{LGAM}
We consider a spherical LG volume, V$_{\rm LG}$, of radius 1.8 Mpc centred on the Milky Way.
Such LG is a large enough volume in order for its halo mass function to be well described by a single power law, for all but the $\sim$ten most massive haloes, as was shown explicitly in \cite{brook14} and also found in other simulations of LG volumes \citep[e.g.][]{gottloeber10,elvis,sawala14}.
We  therefore use a single power law, with slope of $-$0.89, to create a distribution of haloes within the LG, as in \cite{brook14}, with 
\mhalo=$[\rm M_{\rm norm}/N(>$\mhalo$)]^{(1/0.89)}$ where N=0.5,1.5....  is the number of haloes bigger than a given mass and $\rm M_{\rm norm}$ indicates the normalization.

We choose the normalization such that the most massive halo, whether it represents the Milky Way or the Andromeda galaxy, has a virial mass of \mhalo=1.4$\times$10$^{12}$\msun, consistent with mass estimates of the MW \citep{battaglia08,watkins10,Kafle12,bk13}, M31 \citep{corbelli10} and within the values allowed by the timing argument \citep{Kahn59,LiWhite08}.
This results in a LG abundance matching which smoothly joins the large scale surveys \mstar-\mhalo\ relations \citep{guo10,moster10,behroozi13} in a region where the two volumes overlap, i.e. for \mstar$>$10$^{8}$\msun\ (see section~\ref{NORM_LG}).

With this normalization there are 55 dark matter haloes with mass  \mhalo$>$7$\times$10$^{9}$\msun\ within V$_{\rm LG}$, which is the mass above which all haloes should contain stars according to the threshold dictated by reionization \citep{Bullock00,bullock01,Boylan12} and also to recent hydro-simulations \citep{okamoto09,sawala14}. More conservatively, there are 40 haloes more massive than \mhalo$>$1.0$\times$10$^{10}$\msun.

Choosing a different normalization for the most massive halo in the LG will produce different counts in the number of haloes more massive than the reionization limit mass, as in Table 1. 
 
\begin{table}
 \caption{The number of haloes more massive than \mhalo=5, 7 or 10$\times10^9$\msun\ expected in a LG volume V$_{\rm LG}$, considering a power law halo mass function. The count is shown for different normalizations of the most massive halo considered.}
\begin{center}
\begin{tabular}{|ccccc}
\hline
\hline
max \mhalo (10$^{12}$\msun) & 2.0 & 1.4 & 1.0 & 0.8 \\
\hline
$\rm M_{\rm norm}$(10$^{10}$\msun)  & 4.4 & 3.2 & 2.4 & 2.0 \\
N($>$5$\times$10$^{9}$\msun)  & 103 &75 & 56& 47 \\ 
N($>$7$\times$10$^{9}$\msun)  &  76 & 55 & 41& 34\\ 
N($>$10$^{10}$\msun)  &  55 & 40 & 30 & 25 \\ 

\hline
\end{tabular}
\end{center}
\end{table}

\subsection{Abundance matching in the Local Group}\label{AM}
The abundance matching technique constrains  the relationship between the stellar mass and the halo mass of  galaxies \citep{moster10,guo10,behroozi13}. The idea is to match the cumulative number of observed galaxies above a given stellar mass within a given volume, with the number of dark matter haloes within the same cosmologically simulated volume. Such technique is  \textit{independent} of the halo density profile. Abundance matching relations  are generally complete down to a stellar mass of \mstar$\sim10^8$\msun, corresponding to the lower limit of  large scale surveys such as SDSS \citep{baldry08} and GAMA \citep{baldry12}. 
Above this mass, there is relatively little difference in the abundance matching studies \citep{guo10,moster10,behroozi13}, and such differences are insignificant in terms of our study.

Two recent works have extended the abundance matching relation to lower masses using the observed stellar mass function of the LG  \citep{brook14,elvis}. In \cite{brook14}, it was shown that using the average  halo mass function of simulated local groups, which is well described by a power law, implies  a steep relation between \mstar-\mhalo\ in the region 10$^{6.5}$$\lsim$\mstar/\msun$\lsim$10$^8$. The empirically  extended  relation found in \cite{brook14} matches well the extrapolated abundance matching relation of \cite{guo10} and the relation found in \cite{kravtsov10} based on the Milky Way satellites. In \cite{brook14} abundance matching subhaloes are also included within the total halo population, using their peak maximum halo mass value prior to stripping.

The second study \citep{elvis} choses a particular collisionless simulation of the LG  which has a flatter-than-average halo mass function,   allowing the use of a slightly flatter \mstar-\mhalo\ relation when  matching the LG  stellar mass function. Nevertheless, the \cite{elvis} and \cite{brook14} studies are compatible in the mass range 10$^{6.5}$$\lsim$\mstar/\msun$\lsim$10$^8$  to a similar degree as the various abundance matching relations in the mass range \mhalo$\gsim$ 10$^8$\mstar\  \citep{moster10,guo10,behroozi13,moster13}. 

The empirically  extended  relation found in \cite{brook14} and \cite{elvis} implies that galaxies with 10$^{6.5}$$\lsim$\mstar/\msun$\lsim$10$^8$ will all be housed within a narrow halo mass range, 10$^{10}$$\lsim$\mhalo/\msun$\lsim$10$^{10.5}$. 

One assumption behind abundance matching is that every halo contains a galaxy; this may not be true  for the smallest haloes, in which reionization may prevent the collapse of gas and subsequent star formation, leaving dark some haloes with \mhalo$\lsim10^9$\msun\  \citep{Bullock00,Somerville02}.
A correction to the abundance matching relation has been proposed, taking into account this effect, by \citet{sawala14}: by matching galaxies only to some of the smallest haloes, the relation between stellar mass and halo mass becomes flatter. This \textit{adjusted} abundance matching  starts deviating from the usual ones at about \mhalo$<10^{10}$\msun. We emphasize that the LG abundance matching shown in \cite{brook14}  and \cite{elvis} deals with a regime where  \textit{all} haloes are massive enough to be hosts of observed galaxies, therefore no correction is expected in this range where \mstar$\gsim 10^{6.5}$\msun\ and \mhalo$\gsim10^{10}$\msun. 
 
Another factor that is often overlooked is the need to account for the baryon fraction $f_b$. Once abundance matching techniques are used to determine the expected mass of the dark matter halo of each galaxy, the actual  mass of dark matter expected within the  observed galaxy is only (1-$f_b)$$\sim$$0.83$ of that dark matter halo mass. We account for this in our study when  comparing kinematically derived halo masses with dark matter haloes expected from abundance matching techniques.   
Finally,  \citet{sawala14} also suggest that a further adjustment should be made, as they find that halo masses in hydrodynamical simulations can be up to 30\% lower than in dark matter only simulations, i.e. a greater discrepancy than merely accounting for the baryon fraction $f_b$.
We will discuss the consequences of such a reduction in halo mass on our results. 
\subsection{The observational dataset}\label{obs_sigmaV}

We use the luminosities of LG  galaxies, together with their luminosity-weighted average velocity dispersion and half-light radii, as compiled in \cite{kirby14}. The data refer to galaxies with  $10^5$$\lsim$$L_V/L_{\odot}$$\lsim$$2$$\times$$10^8$. Luminosities are converted to stellar masses following \cite{woo08} and \cite{collins14}. 
From the \citet{kirby14} data set, we follow \cite{tollerud14} in excluding NGC 147 and NGC 185, whose relatively large baryonic mass and irregular kinematics make estimates of the dark matter profile highly uncertain, and   follow \cite{gk14} in excluding NGC 6822 for similar reasons.  We have instead added the Sagittarius dwarf irregular galaxy, which has well studied kinematics \citep{cote00} and  adds an extra isolated galaxy to our sample.   
This gives us a sample of 40 dwarf galaxies for which we have de-projected half-light radii $r_{1/2}$ and  dynamical masses within such radii, M$(r_{1/2})$.

It has been shown that stellar velocity dispersions can well constrain the dynamical masses of non-rotating, dispersion supported galaxies at $r_{1/2}$ \citep{wolf10,Walker09}, minimizing the errors introduced by uncertainties of the anisotropy parameter. The mass within the half-light radius of such galaxies reads M$(r_{1/2})=3\langle\sigma^2\rangle r_{1/2}/G$, where $\langle\sigma^2\rangle$ is the luminosity-weighted average of $\sigma^2$ over the whole galaxy  \citep{wolf10}.  The circular velocity is then V$(r_{1/2})=\sqrt{3\langle\sigma^2\rangle}$. This methodology has been shown to work also in non-spherically symmetric systems \citep{thomas11}. 
Some galaxies show significant rotation (Pegasus, WLM, Tucana, and And II): for such galaxies a rotation-corrected velocity for mass estimation is used, that takes into account pressure as well as rotation support in the calculation of M$(r_{1/2})$ \citep{kirby14}.

\subsection{The mass dependent ``DC14" density profile}\label{dc14}

 To model haloes that are flattened by energetic feedback processes, we use the  mass dependent DC14  density profile \citep{DiCintio2014b}, in which   galaxies with  $10^{6.5}$$\lsim$\mstar\ $\lsim$$10^{10}$\msun\ have flatter central densities than the NFW profile \citep{DiCintio2014a}. 

 The balance between the gravitational potential energy of haloes and the energy from the central star forming regions results in a maximum \textit{core formation efficiency} found at  \mstar$\sim$3$\times$10$^{8}$\msun, or equivalently \mstar/\mhalo$\sim0.004$
 \citep[][Brook \& Di Cintio 2015 in prep]{DiCintio2014a}: steeper profiles form in lower mass dwarfs and higher mass  galaxies. For low mass dwarfs, the halo maintains the  NFW profile because there is not enough energy from  supernovae driven outflows to flatten the dark matter cusp, while for high mass galaxies the inner density goes back toward a steep profile because the potential well of the increasingly massive halo is too deep for supernovae feedback to have an effect.

The DC14 profile has been derived using hydrodynamical cosmological simulations \citep{brook12,stinson13} and accounts for the expansion of dark matter haloes as a response to the rapidly changing potential at the center of the galaxy due to the effects of feedback from baryons, particularly due to gas outflows generated in high density star forming regions. 
It takes the form \citep{merritt06}:
\begin{equation}
\rho(r)=\frac{\rho_s}{\left(\frac{r}{r_s}\right)^{\gamma}\left[1 + \left(\frac{r}{r_s}\right)^{\alpha}\right] ^{(\beta-\gamma)/\alpha}}
\label{five}
\end{equation}
\noindent with the two free parameters being the scale radius $r_s$ and the characteristic density $\rho_s$.
The remaining three parameters ($\alpha,\beta,\gamma$), indicating the sharpness of the transition, the outer and the inner slope, respectively, are fully constrained  via the stellar-to-halo mass ratio of a given galaxy, following \cite{DiCintio2014b}:
\begin{equation}
\begin{aligned}
&\alpha= 2.94 - \log_{10}[(10^{X+2.33})^{-1.08}  +  (10^{X+2.33})^{2.29}]\\\
&\beta=4.23+1.34X+0.26X^2\\\
&\gamma= -0.06 + \log_{10}[(10^{X+2.56})^{-0.68}  +  (10^{X+2.56})]\\
\end{aligned}
\label{abg}
\end{equation}
\noindent where $X=\log_{10}$(\mstar/\mhalo).\\
\vspace{-.3cm}

The DC14 profile thus has a range of inner slopes, dependent on the ratio \mstar/\mhalo. 
The scale radius, $r_s$, is connected to the concentration of the halo, defined as $C=R_{\rm vir}/r_s$. 
Such concentration varies with respect to the N-body simulation case \citep{DiCintio2014b}, once $r_s$ has been defined in terms of $r_2$ like in the NFW model:
\begin{equation}
C_{\rm DC14}=(1.0+ 0.00003e^{3.4X})\times C_{\rm NFW}\\\
 \label{c}
\end{equation}
\noindent where $X=\log_{10}$(\mstar/\mhalo) + 4.5.\\
\vspace{-.3cm}

\subsection{Setting Concentrations}\label{concent}
For the NFW profile we tie concentrations to halo masses using the concentration-mass ($C-M$) relation derived within a Planck cosmology \citep{dutton14}. For the DC14 profile, we additionally use \Eq{c} to get the concentration of DC14 haloes once the concentration of NFW haloes has been set. In practice, since we are only dealing with dwarf and intermediate mass galaxies in this study, the two concentrations $C_{\rm DC14}$ and $C_{\rm NFW}$ are the same. 
Of course, scatter in the $C-M$ relation means that individual galaxies may live in more or less concentrated haloes than average. This reasoning, however, cannot be applied to the full set of LG galaxies, as it would be statistically challenging to assume that all of the LG galaxies live in underdense (or overdense) haloes. On average, we do expect that the 40 galaxies studied here should  match the concentration-mass relation, with any scatter not affecting the final trends and conclusions of this work. We acknowledge that satellites are expected to live in haloes which are, on average, more concentrated than isolated ones; we verified that this effect is small \citep{dooley14,gk14} and that it does not influence the results of this study.

Once a density profile and a concentration are set, the halo mass \mhalo\ is the only free parameter left, since ($\alpha$,$\beta$,$\gamma$) in the DC14 profile are constrained by the value of  \mstar/\mhalo\ of each galaxy. 
The assumption that galaxies should follow the empirical mass-concentation relation is central to this study and to the ``too big to fail" problem,  since halo masses are not well constrained  if the concentration is allowed  to be a free parameter.

\section{Results}\label{results}

\subsection{Assigning Local Group galaxies to haloes}\label{match}

 \begin{figure*}
\hspace{-2.cm}
  \includegraphics[width=7.5in]{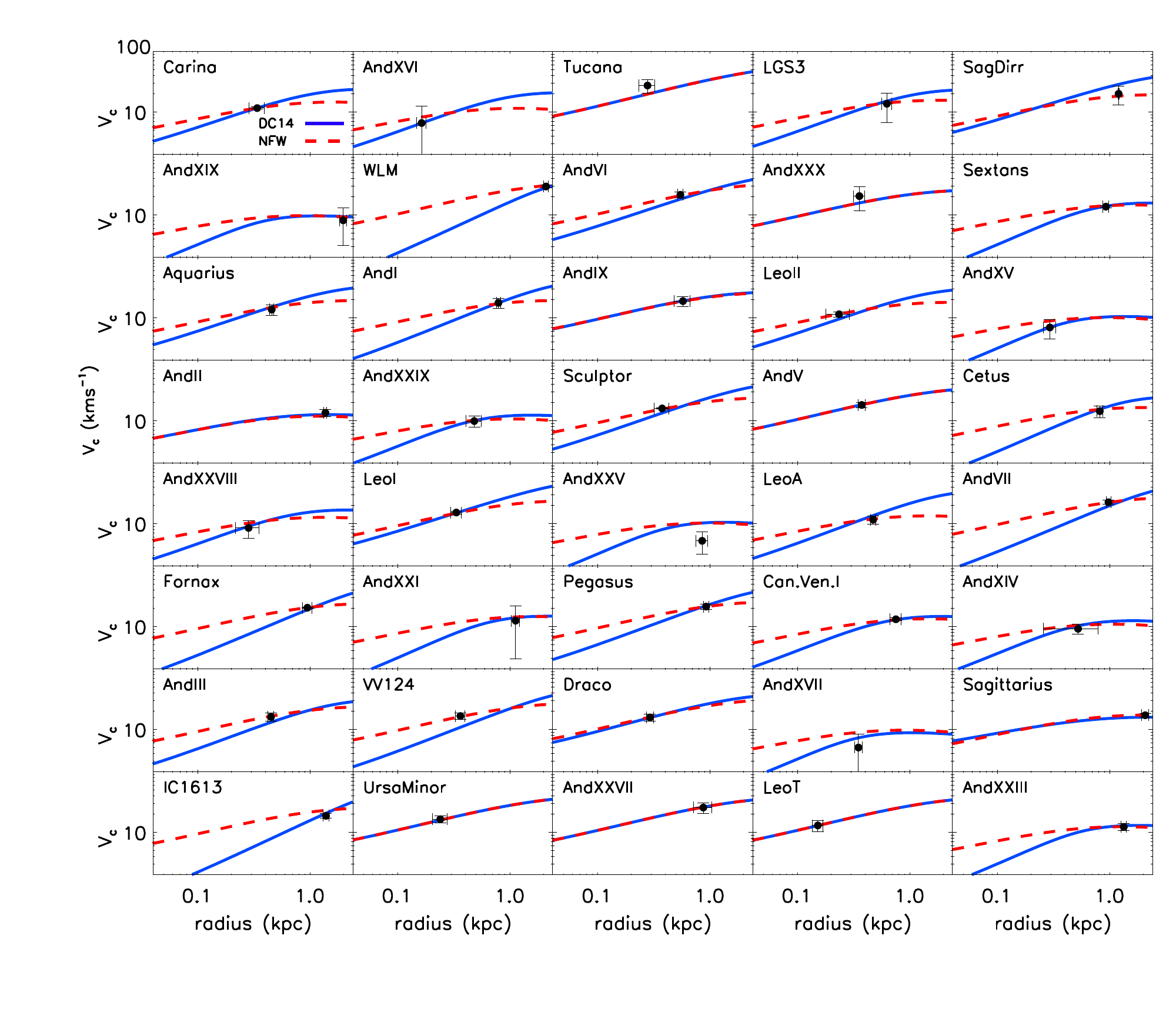}
  \caption{The rotation curves of LG galaxies, assuming an NFW (dashed red) or a DC14 (solid blue) density profile. The black points with error bars are observational data representing the circular velocities  V$(r_{1/2})$, or equivalently the dynamical masses M$(r_{1/2})$, of galaxies within their half-light radii.}
\label{fig:allfits} 
\end{figure*}
  
We find the \mhalo\ that provides the best fit to the observed velocity dispersion of each galaxy in our sample, in terms of circular velocity at the half-light radius, V$(r_{1/2})=\sqrt{3\sigma^2}$. \mhalo\ values are drawn from the available dark matter haloes of the adopted LG halo mass function.

In the process of fitting, we are assuming that the inner region of each galaxy, at its half-light radius, has a density profile that is related to the mass of the halo prior to any tidal stripping from the outer regions. Although satellite and isolated galaxies have undergone though different environmental processes, this is an appropriate assumption, given that the half-light radius of most of the galaxies studied is well within 1 kpc. The \mhalo\ obtained by performing fits to the observed velocity dispersion should therefore be seen as halo masses prior to satellite's infall. \footnote{If, however, a satellite galaxy shows signs of tidal disruption or strong tidal interaction, it is not possible to exclude a priori that the inner region of such galaxy has been affected as well: in such case our methods returns a \mhalo\ which is not necessarily representative of the unstripped, prior-to-infall halo mass. See Section 3.3 and 3.5 for further discussions.}
                       
The ten most massive galaxies of the LG, namely MW, M31, M33, LMC, NGC55, SMC, NGC205, M32, IC10 and NGC3109, are assumed to  be housed within the ten most massive haloes available,  a reasonable assumption given their kinematics and rotation curves.
Therefore the ten most massive \mhalo\ are not available to host the 40 galaxies studied in this work.\footnote{Any scatter in how we assign the 10 galaxies with highest stellar mass to the 10 most massive haloes does not affect our results, so long as these 10 objects are fit within haloes that are above the mass where reionization may cause some haloes to be dark.}

The rotation curves of the best fit halo masses for each galaxy  are shown in Figure~\ref{fig:allfits},  as red dashed lines when assuming an NFW density profile and blue solid lines for the DC14  model. 
Observational data, V$(r_{1/2})=\sqrt{3\sigma^2}$, are shown as black points with error bars and are plotted versus half-light radius. 
Over the entire population, the fitted velocity curves provide a reasonable agreement with the kinematics of observed galaxies, within the observational errors (apart from AndXXV, whose large half-light radius and low velocity dispersion indicates that it lives in a halo smaller than \mhalo$\sim10^8$\msun\ regardless of the profile used, which is the smallest halo considered available; we will come back on AndXXV in \Sec{color}).

In Table 2 we list galaxies stellar masses and derived  halo masses from our  fits, for the NFW profile and the DC14 profile, respectively. For the DC14 profile, we also show the returned values of the transition parameter, outer and inner slopes, ($\alpha$,$\beta$,$\gamma$), corresponding to the best fit haloes. We note that the halo masses recovered in the NFW case are broadly consistent with the halo masses of LG galaxies found in the literature when assuming this profile, i.e. between 10$^8$ and 10$^9$ \msun\ \citep[e.g.][]{penarrubia08,Walker09,gk14}.

\begin{table}\label{t2}
\caption{Compilation of galaxies used in this work. The observational stellar mass and the corresponding halo mass derived by fitting kinematical data using the NFW or the DC14 profile are listed for each galaxy, together with the $(\alpha,\beta,\gamma)$ parameters of the mass dependent profile. Satellite galaxies marked with a $\dag$ do not follow the trends of isolated galaxies, which we interpret as indicating that environmental effects have occurred. In this case both the NFW and the DC14 profile fits would not return the halo infall mass: the quoted values of parameters for the DC14 case may not be valid.}
\begin{tabular}{|l|c|c|c|ccc}
Galaxy & M$_{\rm star}$ & \multicolumn{2}{|c|}{M$_{\rm halo}$    (10$^{8}$$M_\odot$)} \\
Name&  (10$^{6}$$M_\odot$) &NFW  &  DC14&$\alpha$ &$\beta$& $\gamma$ \\

\hline

IC1613 & 102 & 24.4 & 447 & 2.57 & 2.50 & 0.233 \\
WLM & 44.7 & 60.2 & 158 & 2.63 & 2.50 & 0.244 \\
Sagittarius$^\dag$ & 26.9 & 7.00 & 7.00 & 0.854 & 2.85 & 1.09 \\
Fornax & 24.5 & 20.0 & 368 & 2.02 & 2.60 & 0.393 \\
AndVII & 14.8 & 30.2 & 223 & 2.02 & 2.60 & 0.394 \\
VV124 & 10.0 & 29.9 & 237 & 1.81 & 2.67 & 0.507 \\
AndII$^\dag$ & 9.55 & 2.00 & 4.00 & 1.33 & 2.74 & 0.894 \\
AndI & 7.59 & 10.0 & 146 & 1.91 & 2.63 & 0.453 \\
Pegasus & 6.61 & 25.0 & 200 & 1.70 & 2.72 & 0.573 \\
LeoI & 4.90 & 19.9 & 289 & 1.38 & 2.87 & 0.761 \\
AndVI & 3.98 & 59.2 & 190 & 1.48 & 2.82 & 0.700 \\
Sculptor & 3.89 & 20.2 & 152 & 1.58 & 2.77 & 0.643 \\
Cetus & 3.16 & 6.01 & 29.9 & 2.24 & 2.54 & 0.299 \\
LeoA & 2.95 & 3.00 & 80.1 & 1.75 & 2.69 & 0.544 \\
SagDirr & 2.29 & 9.95 & 140 & 1.37 & 2.88 & 0.771 \\
AndIII & 1.82 & 19.8 & 50.2 & 1.74 & 2.70 & 0.548 \\
Aquarius & 1.41 & 9.98 & 61.3 & 1.53 & 2.80 & 0.674 \\
AndXXIII$^\dag$ & 1.26 & 2.30 & 3.99 & 2.65 & 2.50 & 0.255 \\
LeoII & 1.17 & 8.01 & 70.3 & 1.38 & 2.88 & 0.765 \\
AndXXI$^\dag$ & 1.02 & 4.01 & 6.01 & 2.45 & 2.51 & 0.241 \\
Tucana & 0.933 & 404 & 404 & 1.00 & 3.00 & 1.00 \\
Draco & 0.912 & 50.2 & 88.2 & 1.15 & 3.02 & 0.904 \\
Sextans$^\dag$ & 0.851 & 4.00 & 7.02 & 2.30 & 2.53 & 0.278 \\
LGS3 & 0.724 & 5.00 & 19.9 & 1.74 & 2.70 & 0.547 \\
AndXXV$^\dag$ & 0.676 & 1.20 & 2.00 & 2.66 & 2.51 & 0.263 \\
AndXV$^\dag$ & 0.631 & 1.15 & 2.00 & 2.65 & 2.50 & 0.255 \\
AndXIX$^\dag$ & 0.575 & 1.02 & 1.50 & 2.67 & 2.51 & 0.283 \\
UrsaMinor & 0.562 & 101 & 101 & 1.00 & 3.00 & 1.00 \\
AndXVI & 0.525 & 1.80 & 14.0 & 1.76 & 2.69 & 0.538 \\
Carina & 0.513 & 3.99 & 20.0 & 1.58 & 2.77 & 0.644 \\
AndV & 0.398 & 70.3 & 68.8 & 1.00 & 3.00 & 1.00 \\
AndXXVII$^\dag$ & 0.363 & 88.2 & 92.9 & 1.00 & 3.00 & 1.00 \\
AndXVII & 0.309 & 1.01 & 1.14 & 2.62 & 2.50 & 0.241 \\
Can.Ven.I$^\dag$ & 0.309 & 3.00 & 5.00 & 1.99 & 2.61 & 0.411 \\
AndXIV$^\dag$ & 0.302 & 1.50 & 3.20 & 2.19 & 2.56 & 0.320 \\
AndXXVIII$^\dag$ & 0.275 & 2.50 & 6.98 & 1.78 & 2.68 & 0.525 \\
AndXXIX$^\dag$ & 0.234 & 1.40 & 3.00 & 2.10 & 2.58 & 0.358 \\
AndIX & 0.200 & 26.8 & 25.5 & 1.02 & 3.11 & 0.985 \\
LeoT & 0.182 & 90.5 & 90.5 & 1.00 & 3.00 & 1.00 \\
AndXXX & 0.170 & 25.3 & 25.3 & 1.00 & 3.00 & 1.00 \\
 \\\hline
\end{tabular}
\end{table}

While more sophisticated mass modeling techniques have certainly been applied to some of the LG  galaxies  (e.g. \citealt{battaglia08,penarrubia08,Walker09,breddels13}, and references therein), we note that little consensus has been reached so far on the inner slope of dark matter haloes of individual members. 
Indeed, even the best data seem to be unable to  distinguish between cores and cusps in dwarf galaxies (see for example \citealt{battaglia08,Walker11,strigari14}).
Moreover, dynamical Jeans analysis of the radial velocity dispersion profiles of dSphs suffer from the well-known mass-anisotropy degeneracy, which makes difficult to assert the best halo mass value of such galaxies without the knowledge of their anisotropy parameter.

In our study we take another approach, comparing the effects of different density profiles on the population of LG galaxies \textit{as a whole}, using the same method for all galaxies: we will see that specific trends become clear when comparing the resulting \mhalo\ obtained by using NFW versus DC14 profiles, even without using more sophisticated mass models.

Relative to the large range in mass that our sample explores, errors in the mass of individual galaxies are small, allowing us to interpret the trends seen in our ensemble. As mentioned in \Sec{concent}, the advantage of using a full statistical sample of galaxies is that the concentration-mass relation should apply and hold on average, excluding the possibility that all members live in particularly low concentration or low mass haloes as compared with theoretical expectations. 

\subsection{\mstar-\mhalo\ relations for NFW and DC14 profiles}\label{MsMh}
\begin{figure*}
\hspace{-0.35cm}
  \includegraphics[width=3.4in]{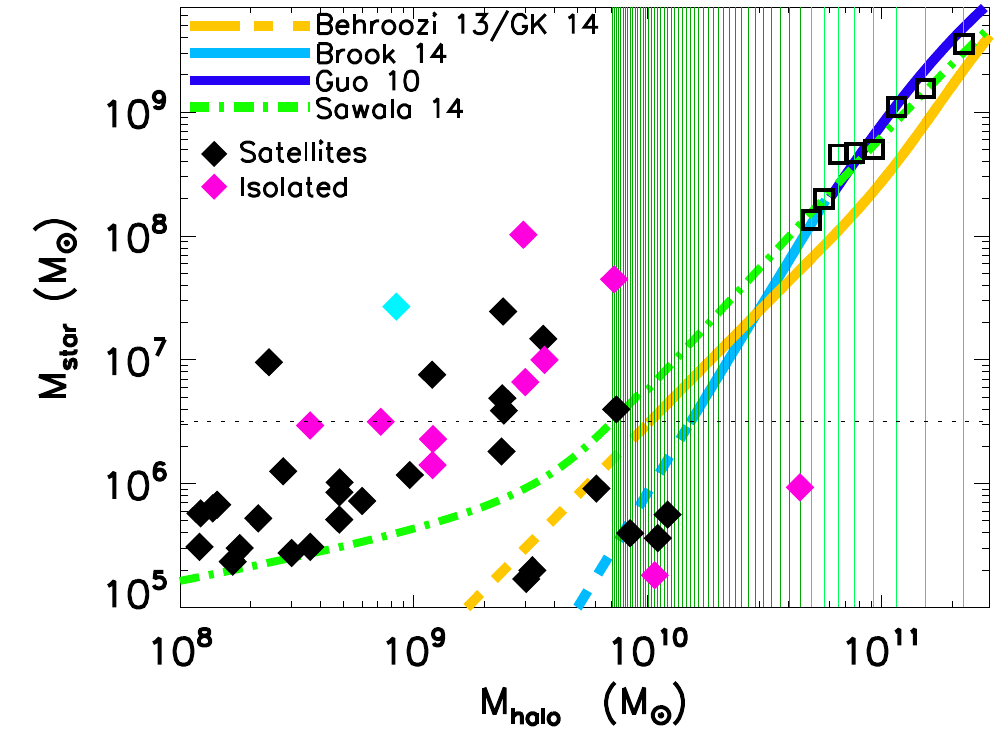}\includegraphics[width=3.4in]{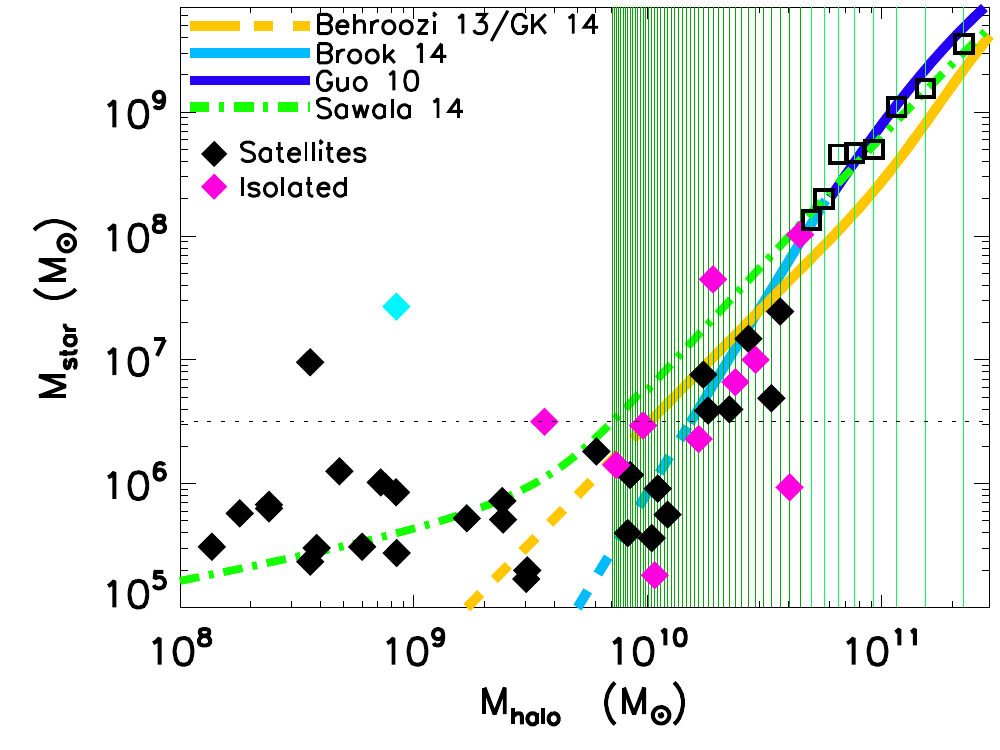}
    \caption{The relation between observed stellar mass and derived halo mass for LG galaxies. The halo mass has been found by fitting kinematical data and assuming two different halo profiles. The results for an NFW profile are shown in the left panel, while the mass dependent DC14 halo profile has been used in the right panel. Satellites and isolated galaxies are shown in different colors, with Sagittarius dwarf irregular, highly affected by tides, shown in cyan. Several abundance matching predictions are indicated, in particular the \citet{brook14} one has been constrained using the LG mass function, and it is shown as dashed line below the observational completeness limit of the LG.}
\label{fig:NFW}
\end{figure*}

In Figure~\ref{fig:NFW} we  compare results for the ensemble of LG galaxies obtained by assuming either an NFW profile (left panel) or a DC14 one (right panel). The green vertical lines are the 55 dark matter haloes, from our halo mass function, with \mhalo$>$ 7$\times$10$^{9}$\msun: they represent the ``too big to fail" haloes, as in this region we expect all haloes to have stars.

The \mhalo\ values from our fits are shown as diamonds and plotted versus the \mstar\ of each galaxy. 
Black diamonds are satellite galaxies of the Milky Way and Andromeda galaxy, while magenta diamonds are isolated galaxies.
Sagittarius dwarf spheroidal is indicated in cyan, as its stellar mass and kinematics (and therefore derived halo mass) have been significantly affected by its tidal disruption. 
We have accounted for the 17\% reduction in halo mass due to the universal baryon fraction $f_b$ when comparing with abundance matching expectations from collisionless simulations, by actually plotting M$_{\rm halo}$/(1-$f_b$), where M$_{\rm halo}$ is the halo mass derived from the kinematic fits.

Several abundance matching relations are superimposed on the plot \citep{guo10,brook14,elvis,sawala14}. 
As stated in \Sec{LGAM}, when using the fiducial LG  mass normalization of max(\mhalo)$=$1.4$\times$10$^{12}$\msun, the most massive galaxies (not subject of this study, and indicated as open squares) lie on the \mstar-\mhalo\ plane in a way that joins nicely the abundance matching relations constrained by large scale surveys \citep{guo10}.
The horizontal dotted line shows the approximate  observational completeness limit of the LG, within a sphere of $\sim1.8$ Mpc of radius from the Milky Way: the census of galaxies with stellar masses lower than $\sim$3-5$\times$10$^{6}$\msun\ is likely incomplete in this volume \citep{Tollerud08}.

The difference between the two derived  \mstar-\mhalo\ distributions is striking.
For the NFW profile, the observed kinematics imply that the galaxies are well fit by haloes with masses $10^{8}$$\lsim$\mhalo/\msun$\lsim$$10^{10}$: only one galaxy from our sample fits to a halo which is more massive than $10^{10}$\msun, in disagreement with abundance matching predictions \citep{brook14}.
Using the NFW profile to describe the kinematics of LG members highlights the ``too big to fail" problem  in the LG  \citep{gk14,kirby14}: almost none of the most massive haloes, indicated as vertical green lines, has been assigned to a galaxy. 
The \textit{darkness} of the haloes with mass  \mhalo$>$ 7$\times$10$^{9}$\msun\ can not be explained simply by invoking reionization processes, as these haloes are more massive than the reionization mass threshold \citep{Boylan12}. 

The preferred values of \mhalo\ resulting from an NFW model are in disagreement with abundance matching relations even after taking into account the reduction of dark matter mass by up to a $30\%$ caused by baryonic effects and the fact that some of the smallest haloes remain dark \citep[][]{sawala14}.

A different picture appears for the DC14 profile, shown in the right panel of Figure~\ref{fig:NFW}.
Now the most massive dwarfs, those with \mstar$\gsim$$3$$-$$5$$\times$$10^6$\,\msun, all fit in haloes more massive than \mhalo$\sim$$10^{10}$\msun\ (apart from two outliers far from equilibrium, namely Sagittarius dSph, which is being disrupted, and AndII, which shows signs of a merger, \citealt{amorisco14}): the  distribution of preferred halo masses is therefore shifted toward the right side of the plot.
This can be easily understood in terms of halo profiles: galaxies with a \mstar/\mhalo\ ratio higher than 0.0001 develop the minimum amount of energy, from stellar feedback, required for their profile to be shallower than NFW. In this way a galaxy like Fornax, for example, with \mstar$\sim$$2$$\times$$10^7$\msun, will be well fit by an NFW halo with \mhalo$\sim$$2$$\times$$10^9$\msun\ or by a more massive  DC14 halo with \mhalo$\sim$$3$$\times$$10^{10}$\msun\ and inner slope $\gamma$$\sim$$-0.39$. 

Moving the distribution of \mstar-\mhalo\ to the right has two consequences: firstly, the number of ``too big to fail" haloes is considerably reduced, as we are assigning galaxies to haloes more massive than \mhalo$\sim7\times10^9$\msun; secondly, the distribution is now in agreement with the abundance matching predictions of \citet{brook14} and \citet{elvis}, down to the observational completeness limit of the LG.

Below the completeness limit of \mstar$\sim$$3$$-$$5$$\times$$10^6$\,\msun, many galaxies still prefer to live within less massive haloes than an {\it extrapolation} of abundance matching would predict, even when applying the DC14 profile. A comparison with the proposed abundance matching of \citet{sawala14} seems to provide a good agreement with the kinematic of such low mass galaxies, when the DC14 profile is assumed. 

However, all the dwarf galaxies in this low halo mass region are satellites of either the Milky Way or Andromeda: the fact that no isolated galaxy is found within such low mass haloes, \mhalo$\lsim3$$\times$$10^9$\,\msun, suggests that environmental effects are in place. Indeed, for most of the satellites in this region, signs of tidal interaction have been invoked in the literature \citep{okamoto12,battaglia11,battaglia12}, and we will name and discuss these objects in section~\ref{environment}. 
Numerical simulations suggest that  environmental effects may be  important even in the inner region of galaxies once baryonic physics have been taken into account, since the presence of a baryonic disk can enhances tidal effects \citep{brooks13,arraki13}. Tidal effects are even more important in those satellites that formed a core at early times, before infall into the main host \citep{penarrubia10,zolotov12,madau14}. 
Tides are therefore a possible mechanism to reduce the masses of such galaxies \citep{collins14}.

Our analysis therefore supports the notion that a combination of dark matter halo expansion due to baryonic effects and enhanced environmental processes can explain the kinematics of LG  galaxies. Satellite galaxies living in the low halo mass region have likely been placed there because their kinematic has been affected by tides, and they would have had a higher halo mass otherwise \citep{collins14}, bringing them in agreement with the \citet{brook14} and \citet{elvis} abundance matching (and their extrapolation). On the other side, all the isolated galaxies match  such relations when the DC14 model is assumed. 

\subsection{Environmental Effects?}\label{environment}
\begin{figure}
\hspace{-0.35cm}
  \includegraphics[width=3.4in]{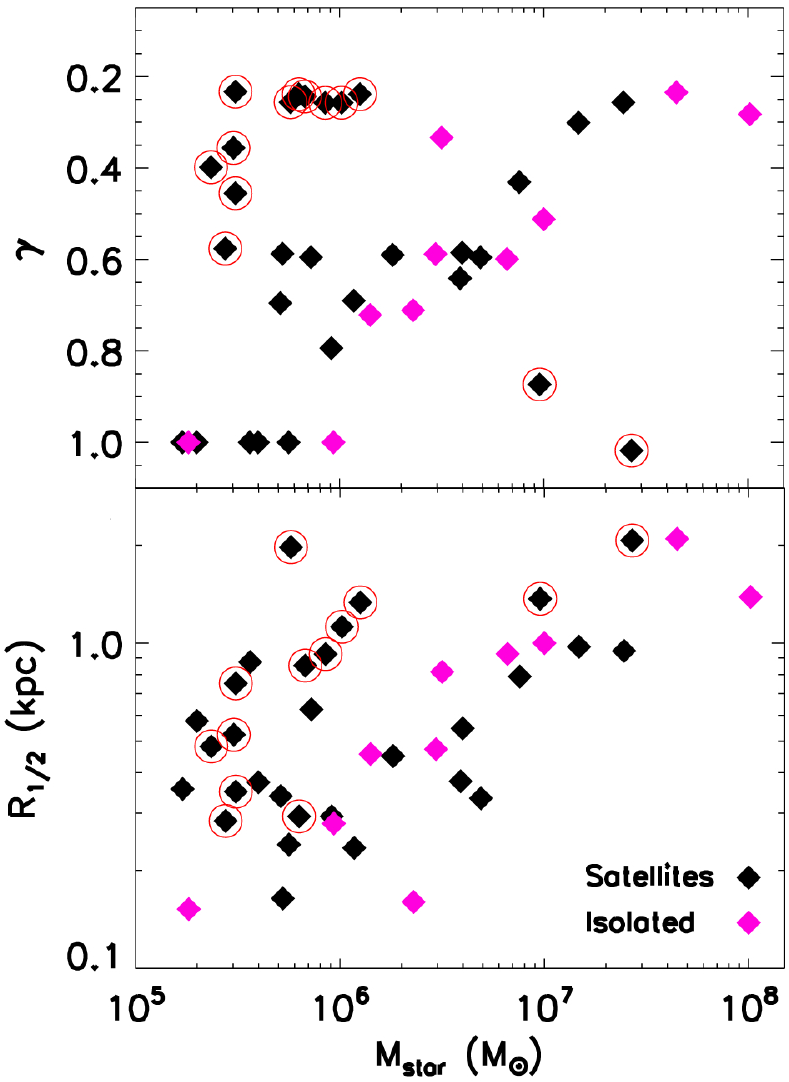}
   \caption{The upper panel shows the derived inner slope, $\gamma$ from DC14 profile, against \mstar\ for isolated (magenta) and satellite (black) LG galaxies. Satellites whose best fit halo mass is low  compared to abundance matching expectations (\mhalo $<$10$^9$\msun) are marked with red circles: they  do not follow the same trend of decreasing $\gamma$ (cored profile) for increasing \mstar\ as isolated galaxies. The lower panel shows the half-light radii versus \mstar, using the same symbols.  The satellites living in low halo masses have relatively high half-light radii compared to the isolated galaxies of similar \mstar, a further indication that they may have been tidally affected.} 
\label{MsRhalf}
\end{figure}

To test further the notion that environmental effects may play a role in determining the kinematics of  satellite galaxies that are assigned to low mass haloes, we show the galaxies in the plane of \mstar\ versus inner slope $\gamma$ in the top panel of Fig~\ref{MsRhalf}, where satellite galaxies are indicated in black  and isolated objects in magenta. Galaxies whose best fit halo is less than 10$^9$\msun\, in the DC14 case, are marked with red circles, and it is evident how they do not follow the same relation of increasingly cored profile for increasing stellar mass followed instead by isolated galaxies. 
The trend of lower $\gamma$ for higher \mstar\ is in very good agreement with previous results from \citet{zolotov12} who studied satellite galaxies within hydrodynamical simulations.

We then plot, in the lower panel of Fig~\ref{MsRhalf},  the half-light radius \rhalf\ versus  stellar mass of LG galaxies. The satellites that we have interpreted as being environmentally affected are again marked with red circles; they tend to have larger \rhalf\ values at given \mstar, compared to the isolated sample, as expected if tidally effects have been in place.
If environmental effects have influenced this sub-sample of satellite galaxies, then we emphasise that both the NFW and the DC14 profile would not return the satellite halo mass at its peak, i.e. at infall time, as it would be desirable, but rather the z=0 halo mass. This would invalidate the values given in Table~\ref{t2} for these satellite galaxies, which we indicate with a $\dag$. In particular, the inner slopes shown for those particular satellite galaxies should not be considered a prediction of this study. 
Indeed the DC14 density profile, parametrized as a function of \mstar/\mhalo, accounts for the impact of outflows of gas within galaxies but has not been designed to include tidal effects. 

It is possible however that cores have formed in some of these satellite galaxies at earlier times, when the energetic requirement for core formation was lower given the lower halo mass \citep{penarrubia12,amorisco14,madau14}, and that such galaxies have further evolved and interacted with the main hosts, causing tides to play an important role in their dynamics \citep{arraki13,brooks13}. 
 
 Two satellite galaxies appear in the upper, left region of  Figure~\ref{fig:NFW} . The first, Sagittarius dwarf spheroidal \citep{ibata94}, is currently interacting with the Milky Way's disk and has been severely stripped and disrupted; the second one, the Andromeda II satellite of M31, has been reported to be the remnant of a merger between two dwarf galaxies \citep{amorisco14}, which may have had a strong impact on its kinematics.

Excluding  Sagittarius, three Milky Way satellites are found not to lie on the abundance matching relation, namely Sextans, Carina and Canes Venatici I, with some of them living in a region where shallow profiles are preferred. The irregular, distorted shape of Canes Venatici I could indicate that it is  suffering from tidal effects \citep{okamoto12}, although it is one of the most distant satellites of our Galaxy. It has been argued that the young stellar population of this galaxy, having a small velocity dispersion, could reflect the presence of a cored profile \citep{ibata06}. 
Sextans shows remarkable velocity gradients that could have been caused by its tidal disruption from the Milky Way \citep{battaglia11} and its observed stellar clumps are in agreement with a cored dark matter halo \citep{lora13}.

Finally, Carina's spatial extent of member stars \citep{kuhn96} and observed tidal debris \citep{battaglia12} provide evidence of tidal interactions, although such tidal features have been  recently found to be weaker than previously thought \citep{monigal14}.

The remaining galaxies which prefer to live in haloes smaller than \mhalo$\sim$3$\times$$10^9$\msun\ are satellite galaxies of Andromeda.
AndXIV, AndXV, AndXVI, AndXIX, AndXXI and AndXXV are extreme outliers, given their high half-light radius and low velocity dispersion, and has been suggested \citep{collins14} that tides may explain their associated low masses.
\citet{collins14} points out that the fact that more Andromeda satellites fit to  low mass haloes than Milky Way satellites do, may indicate that these systems have been, in general, more strongly affected by tides.

Moreover, the z=0 halo mass for AndXXV must be even lower than \mhalo$\sim10^8$\msun, given that we could not fit its kinematic with any of the available haloes from our halo mass function, whose lowest considered  halo is \mhalo$\sim10^8$\msun: as pointed out in \citet{collins14}, with such a low halo mass this galaxy should not have formed any stars, unless its mass was significantly higher in the past, prior to being accreted.

Support for the notion that satellites assigned to low mass halos by our model have been affected by the environment comes from the work of \citet{zolotov12}, in which it was shown that satellites in hydrodynamical simulations may have their velocities reduced even in the inner 1kpc.

\subsection{On the normalization of the Local Group mass}\label{NORM_LG}

A degree of uncertainty exists in the mass estimates of the MW \citep[][and references therein]{battaglia08,bk13,watkins10,Kafle12,Kafle14}, M31 \citep[e.g.][]{corbelli10} and of the LG  \citep{klypin02,LiWhite08,vandermarel12,gonzales14,penarrubia14,diaz14}: we explore here how this affects our results. Our fiducial model assumes that the most massive galaxy, whether it is the MW or M31, has a mass of \mhalo$=$$1.4$$\times$10$^{12}$\msun, close to the most favored observationally determined values.

\begin{figure}
\hspace{-0.35cm}
  \includegraphics[width=3.4in]{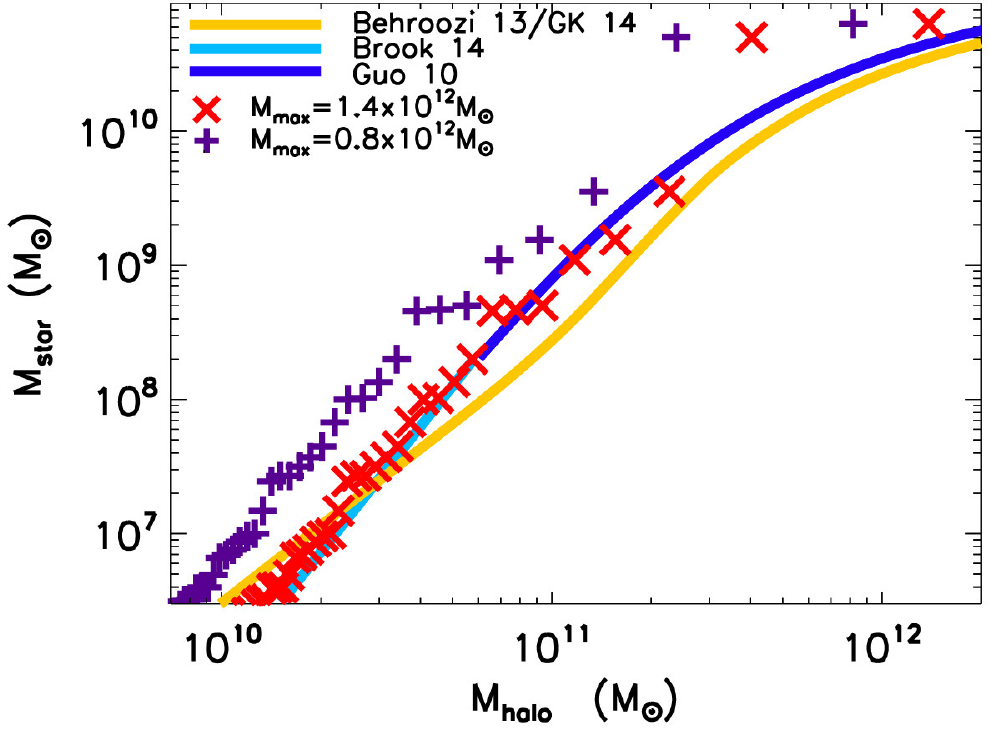}
   \caption{The abundance matching in the LG, derived by ordering galaxies by stellar mass and matching them to a power law halo mass function with two different normalizations: a maximum halo mass of  \mhalo$=$$1.4$$\times$10$^{12}$\msun (red crosses) or a lower maximum halo mass of  \mhalo$=$$0.8$$\times$10$^{12}$\msun\ (violet plus) have been used. The distribution of red crosses is equivalent to the \citet{brook14} slope, cyan line here and in Figure 2, which has indeed been derived using such fiducial maximum halo mass: in this case both the slope \textit{and} the normalization of LG  galaxies are in agreement with abundance matching results of large galaxy surveys \citep{guo10,behroozi13}. A lower normalization  would instead result in the ensemble of LG  galaxies being matched to systematically lower mass haloes than predicted from large surveys, which are well constrained for \mstar$\gsim10^8$\msun.} 
\label{AMLG}
\end{figure}

In Figure~\ref{AMLG} we show the abundance matching relations that arise using such a fiducial maximum halo mass, as well as a normalization with maximum mass of \mhalo$=0.8\times10^{12}$\msun, which is around the lowest estimate of the Milky Way and Andromeda mass. We  compare the results with large scale surveys abundance matching.

The red crosses are LG  galaxies matched to corresponding haloes\footnote{We are in a region where all haloes are expected to host galaxies, therefore a one-to-one matching is appropriate.} for the fiducial model: such matching results in the already shown (Figure 2) slope of \citet{brook14}, indicated as cyan solid line.  
Interestingly, when using this fiducial model, the highest mass LG  galaxies are matched to haloes in such a way that the resulting relation  between \mstar\ and \mhalo\ is in agreement with large volume abundance matchings \citep[][shown as dark blue and yellow line, respectively]{guo10,behroozi13}  in the mass region where the two different volumes overlap, and matches their extension to lower masses  \citep{brook14,elvis}.

By contrast, choosing the lower mass normalization results in LG galaxies  being assigned to systematically less massive haloes than found in abundance matching studies of large scale surveys \cite[e.g.][]{guo10,moster13}, in a region where the two volumes overlap, as can be seen from the violet plus signs compared to the dark blue line in Figure~\ref{AMLG}.
Moreover, a maximum halo mass as low as \mhalo$=0.8\times10^{12}$\msun, although allowed by current  estimates of the MW and M31 mass \citep{battaglia08,Kafle14,corbelli10}, will result in tension with the lowest LG mass estimates from timing arguments \citep{LiWhite08}.

\begin{figure*}
\hspace{-0.35cm}
  \includegraphics[width=3.4in]{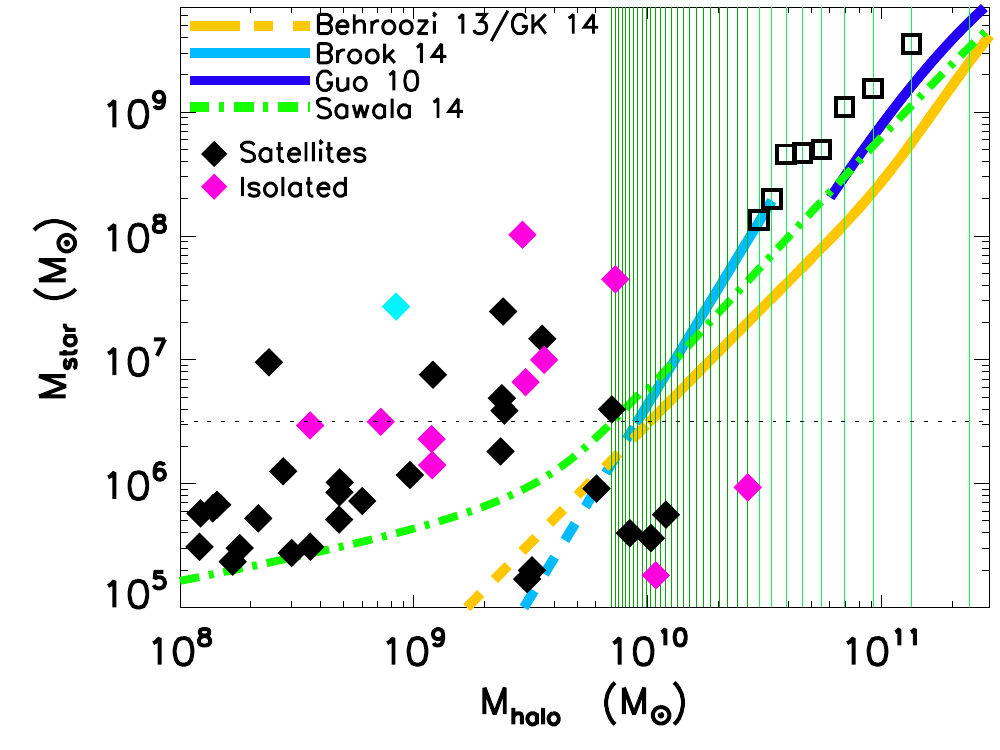}\includegraphics[width=3.4in]{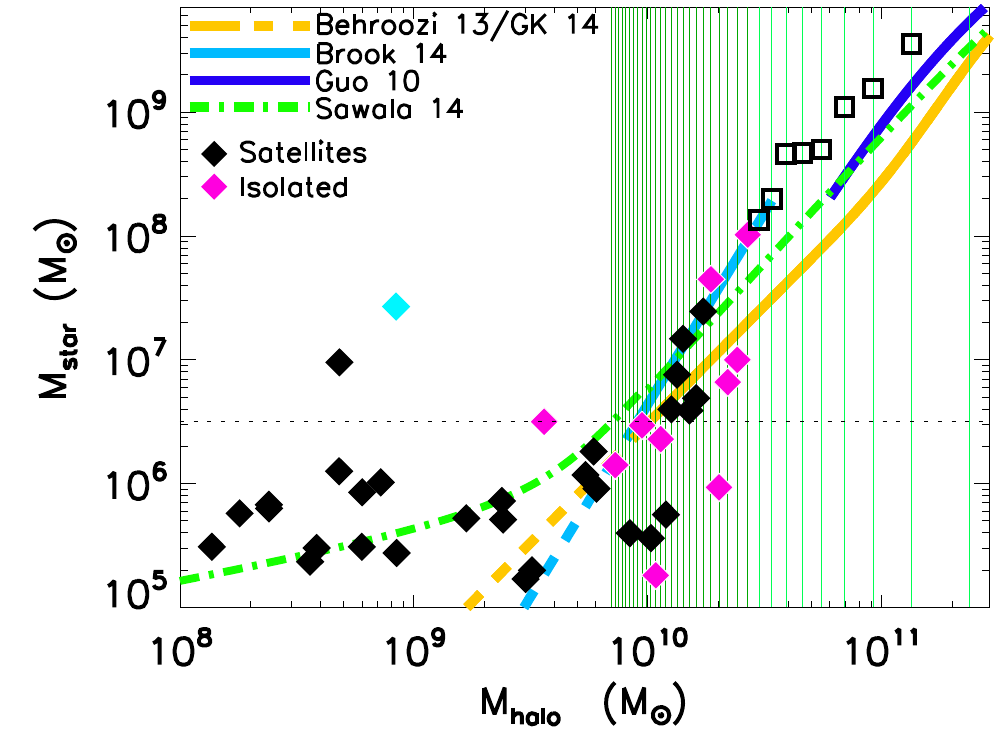}
    \caption{Same as in Figure~\ref{fig:NFW}, but for a lower normalization of the LG mass. The most massive galaxy in this case has a halo mass of \mhalo$=$0.8$\times$10$^{12}$\msun, which correspond to the lowest mass estimates of MW and M31. The results for an NFW profile are shown in the left panel, while the mass dependent DC14 halo profile has been used in the right panel. Given the change in halo mass function normalization, the \citet{brook14} abundance matching has been shifted accordingly, and is equivalent to the distribution of violet plus signs of Figure~\ref{AMLG}.}
\label{fig:NFW_lowermass}
\end{figure*}

Nevertheless, to show that our main results are only marginally affected by the chosen normalization of the LG, in Figure~\ref{fig:NFW_lowermass} we repeat the analysis using the low mass normalization of the halo mass function, max(\mhalo)=$0.8$$\times$10$^{12}$\msun. We remind that we fit kinematics of the sample of LG galaxies with the most appropriate halo drawn from this halo mass function. The best fit haloes when assuming the NFW profile (left panel) have almost not changed compared to the results shown in Figure~\ref{fig:NFW}. 
What has changed is the number of ``too big to fail" haloes, shown as green lines, which we discuss below. 
 
For the DC14 profile, a lack of availability of appropriate mass haloes from the halo mass function means that several of the highest mass galaxies are matched to haloes with lower mass than was the case in Figure~\ref{fig:NFW_lowermass}, meaning haloes with lower mass than their kinematic fits warrant. One way to look at this is that the amount of expansion predicted in the DC14 profile matches the kinematics of the LG galaxies for the fiducial normalization, but a lower normalization would be best fit by models with slightly less halo expansion.  
Regardless, the main differences between the NFW and DC14 distributions are retained, confirming our finding that expanded haloes are needed to match abundance matching predictions in the LG.

The degree to which the ``too big to fail'' is solved depends on the normalization of the LG.  As from Table 1, there are 55 haloes more massive than \mhalo$\sim7\times10^9$\msun\ for our fiducial normalization and 34 for the lowest mass normalization, shown as vertical green lines in  Figure~\ref{fig:NFW} and Figure~\ref{fig:NFW_lowermass}, respectively. The completeness limit of the LG must be considered  when counting the ``too big to fail'' haloes that have not been assigned to  a galaxy, as uncertainties remain about the exact number of galaxies with  \mstar$<$$3$$\times$$10^6$\msun, and some of them may be expected to match haloes with \mhalo$\gsim7\times10^9$\msun.

Considering therefore the completeness limit of LG surveys, for our fiducial normalization, there are $\sim28$ haloes more massive than \mhalo$\sim1.5\times10^{10}$\msun, i.e. above the point where the \citet{brook14} abundance matching in Figure~\ref{fig:NFW} crosses the line where \mstar$=$$3$$\times$$10^6$\msun: 10 of such haloes are assigned to the LG galaxies studied here if a DC14 profile is used, and another 10 are expected to host the most massive LG members (MW, M31 etc., indicated as open squares in Figure~\ref{fig:NFW}).
Asides from the 10 most massive galaxies, there are another 8 galaxies (NGC 185, NGC 147,  IC 5152, Sextans B, Sextans A, And XXXI, And XXXII, UKS2323-326) within the completeness limit, i.e. with stellar masses \mstar$>3\times10^6$\msun,  not included in our sample because their current kinematic information made it difficult to determine their halo masses. 
Assigning these 8 galaxies to the remaining 8 massive haloes would completely solve the ``too big to fail" problem in a region where the census of LG galaxies is complete.\footnote{In the following discussion we assume that these 8 galaxies live into massive haloes, \mhalo$\gsim7\times10^9$\msun, for both the NFW and DC14 case.}

The situation is different for an NFW profile, in which none of the studied LG galaxies is assigned to haloes more massive than \mhalo$\sim10^{10}$\msun, leaving $\sim9$ massive failures above the completeness limit of the LG.

Below the completeness limit, counting all the haloes more massive than \mhalo$\sim7\times10^9$\msun, the DC14 profile results in 17 haloes unaccounted for versus 30 in the NFW case.
Recall, however, that several of the lowest mass (\mstar$\lsim3\times10^6$\msun) satellite galaxies have likely been affected by environmental processes, so that their estimated halo mass is actually a lower limit of their virial mass: assuming that some of these galaxies have a mass higher than what suggested by their z=0 kinematics will further help in bringing the number of observed galaxies in agreement with theoretical predictions, once the DC14 profile is assumed. 

Finally, we want to highlight the  case of Tucana: it is an extremely low luminosity galaxy that fits nevertheless to a  halo \mhalo$\gsim10^{10}$\msun, regardless of the density profile used. 
The census of LG galaxies within 1.8 Mpc of the Milky Way is complete only down to  \mstar$>3-5\times10^6$\msun\ \citep{Koposov08,Tollerud08} and future deep surveys will hopefully increase the agreement between theoretical expectations and observations, if a few new faint galaxies with similarly high velocity dispersions as Tucana are discovered.

Using the lower normalization, as in Figure~\ref{fig:NFW_lowermass}, will result in the abundance matching (now shifted to the left to reflect the average slope and normalization of the violet signs in Figure~\ref{AMLG}) to hit the observational completeness limit at \mhalo$\sim8.8\times10^{9}$\msun, with again 28 haloes more massive than this value.
In the DC14 case, every halo more massive than \mhalo$\sim7\times10^9$\msun\ (34 in total) has been assigned to a galaxy, while in the NFW case we are still left with 9 of such haloes not forming any stars, all of them lying in a region where the LG is observationally complete

\subsection{Star formation efficiency in isolated Local Group galaxies}\label{sfe}

Abundance matching relations from large scale surveys have shown that the galaxy formation efficiency, \mstar/\mhalo,  decreases sharply as \mhalo\ decreases for \mhalo$\lsim10^{11.5}$\msun: below this mass the relation can be approximated by a power law, \mstar$\sim$\mhalo$^{\alpha}$, whose slope  depends on the observed faint end of the luminosity function, leading to values of $\alpha\sim3$ \citep{guo10} and $\alpha\sim2.4$ \citep{moster13}.
LG  abundance matching has shown that the relation extends to significantly lower stellar mass, \mstar$\gsim$$10^{6.5}$\msun, providing values of $\alpha\sim3$ \citep{brook14} and $\alpha\sim1.9$ \citep{elvis}. \footnote{\cite{kravtsov10} infers  a similar relation, with slope $\alpha$$=$2.5, from the MW satellite luminosity function. However, this estimate has larger uncertainties than the afore mentioned studies due to the less well constrained halo mass function of an individual MW dark matter halo, and due to the considered mass range in which some halo may be dark due to reionization.} 

Although the variation between these results is not insignificant, they all indicate  that \mstar\ decreases sharply as \mhalo\ decreases, i.e. the efficiency of a galaxy in converting baryons into stars decreases with decreasing \mhalo, such that dwarf galaxies are the objects with the lowest star formation efficiency in the Universe.

\begin{figure}
\hspace{-0.35cm}
  \includegraphics[width=3.4in]{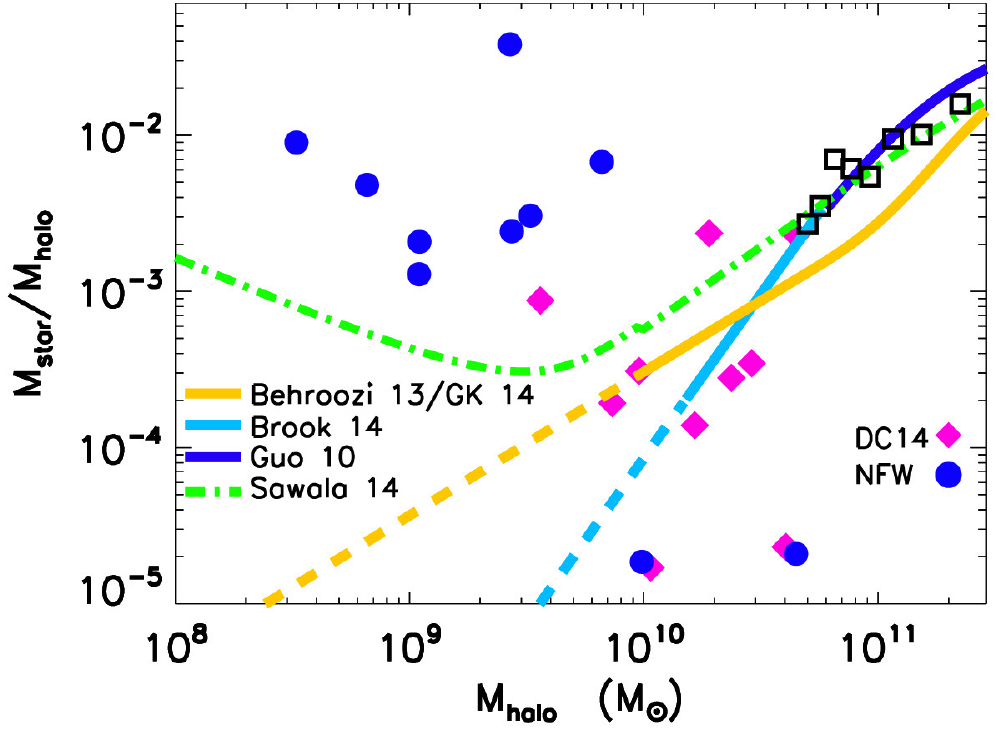}
   \caption{The galaxy formation efficiency, \mstar/\mhalo, as a function of \mhalo\ for isolated LG galaxies. The best fit halo mass has been derived  using an NFW (blue circles) or a DC14 (magenta diamonds) profile. The solid and dashed lines represent different abundance matching prescriptions.}
\label{fig:sfe_fig}
\end{figure}

An interesting aspect of our analysis is reported in Figure~\ref{fig:sfe_fig}, in which the star formation efficiency, \mstar/\mhalo, is shown as a function of \mhalo\ for the isolated LG galaxies from our sample. 
Best fit halo masses  are derived assuming an NFW profile (blue circles) or DC14 profile  (magenta diamonds). The solid and dashed lines are predictions from different abundance matching relations.
Regardless of LG  mass, the NFW model places galaxies with $10^6$$\lsim$\mstar/\msun$\lsim$$10^8$ within haloes of $10^8$$\lsim$\mhalo/\msun$\lsim$5$\times$10$^9$, implying a reversal of the steep trend of decreasing star formation efficiency, also observed if the \textit{adjusted} abundance matching \citep{sawala14} is applied: several of these galaxies are assigned to such small haloes that their efficiency is comparable to the one of objects 2-3 orders of magnitude more massive.

A reversal of the trend is difficult to explain within our current theoretical framework of galaxy formation. Rather, one may expect that at lower galaxy masses, the combined effects of feedback and reionization would further suppress star formation efficiency.

Indeed, in low-mass systems, both winds from supernovae \citep{oppen08} and the presence of a UV background can further reduce the amount of gas that cools at the center of a halo \citep{Navarro97UV}: reionization, in particular, completely prevents the formation of galaxies within several of the smallest haloes \citep{Bullock00}.
Hydrodynamical simulations have shown that  the steep declining in \mstar/\mhalo\ holds at low masses, with ultrafaint dwarfs of stellar masses $10^4\leqslant$\mstar/\msun$\leqslant10^5$ expected to inhabit haloes of a few times in $10^9$\msun\ \citep{brook12,munshi13,Shen14,hopkins14}. Semi-analytic models also predict  that star formation efficiency will continue to decrease as halo mass decreases \citep{Benson02,guo11}.

To explain the star formation efficiency of LG dwarf galaxies inferred using the NFW model, one would require a physical mechanism that leads such low mass haloes, \mhalo$\lsim10^9$\msun, to have the same efficiency, in terms of \mstar/\mhalo, as haloes with mass \mhalo$\sim$$10^{11}$\msun. This reasoning is specially important for isolated LG  members, in which environmental effects can not be invoked to reduce their halo masses.

By contrast, when halo masses are derived assuming a DC14 profile, the efficiency of the isolated dwarfs matches the relation between \mstar/\mhalo\ and \mhalo\ implied by the abundance matching of \citet{brook14}, \citet{gk14} and their extrapolations below \mstar$\sim$3$\times$10$^{6}$\msun, as well as current cosmological simulations.
While some scatter is observed, and may be expected at such low halo masses due to the variety of star formations histories \citep{weisz14} and merging histories of such systems, the average trend remains clearly a decline in \mstar/\mhalo\ as we lower the halo mass, in agreement with both large scale and  LG abundance matchings.

From another perspective, if one {\it assumes} that the slope of \mstar/\mhalo\ as a function of \mhalo\ extends to the mass range of LG  dwarf galaxies, one would {\it require} expanded haloes to explain the  kinematics of such galaxies.
 \\\\
\subsection{The cusp/core space of Local Group galaxies}\label{color}

The cusp/core transformation within galaxies is predicted to depend on the ratio of stellar-to-halo mass, and it is expected  if 0.0001$\lsim$\mstar/\mhalo$\lsim$0.03. 

\begin{figure*}
\hspace{-0.35cm}
  \includegraphics[width=5.2in]{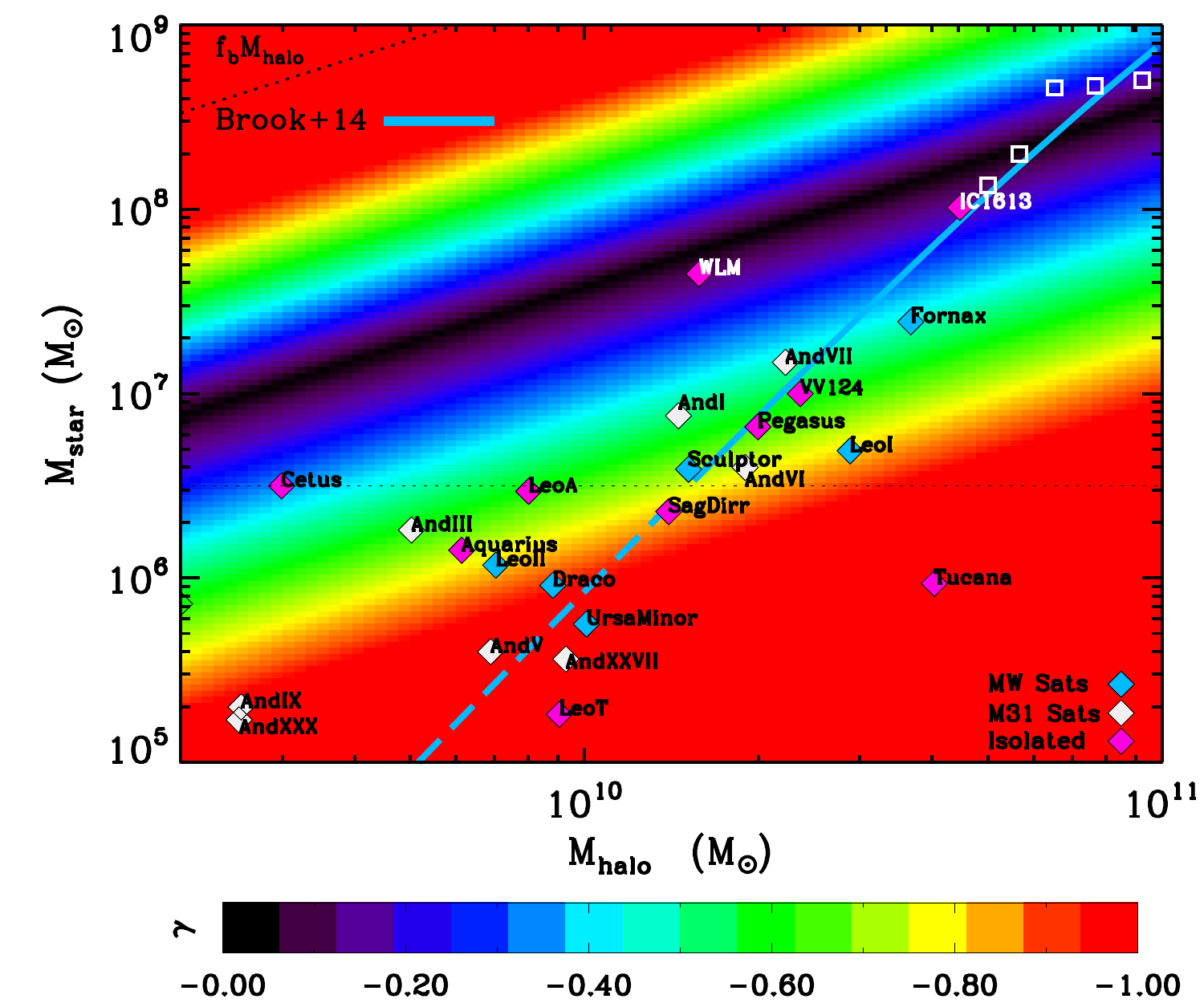}
    \caption{The stellar mass versus halo mass region populated by LG  galaxies, assuming that they follow the mass dependent density profile DC14 proposed in \citet{DiCintio2014b}. The cyan solid line is the abundance matching prediction from \citet{brook14}, which is well constrained down to the completeness limit of current surveys, i.e. down to a stellar mass of $10^{6.5}$\msun; below such limit the relation has been linearly extrapolated, and is shown as dashed cyan line. The galaxies are color coded according to their membership: Milky Way's satellites in cyan, Andromeda's satellites in white and isolated galaxies in magenta. The background image is colored according to the inner slope of dark matter profile expected for any given \mstar/\mhalo\ value: cuspy haloes ($\gamma=-1.0$) live in the red region, while the cored-most ones ($\gamma=0.0$) are found in the dark area. It is immediate to see which LG member prefers cusps or cored profiles.\hspace{.2cm}
    (A color version of this figure is available in the online journal.)}
\label{fig:color_panel}
\end{figure*}

In Figure~\ref{fig:color_panel} we provide a different perspective of our findings, showing the \mstar-\mhalo\ results from the right panel of Figure~\ref{fig:NFW}  imposed on the ``cusp/core" space, colored by the value of the inner slope $\gamma$  according to the DC14 profile \footnote{Note that here, since we are mainly interested in a comparison with the cusp/core space rather than abundance matching prediction, we did not increase the halo mass for the baryon fraction, and the \mhalo\ values shown are exactly those listed in Table 2.}. We have  excluded the satellite galaxies that we believe may have been influenced by tides, those marked $\dag$ in table~\ref{t2}, as the inner slope results may not be appropriate in such cases. 

Galaxies are explicitly named and color coded according to their membership, with Milky Way's satellites in cyan, Andromeda's satellites in white and isolated galaxies in magenta. The \citet{brook14} abundance matching is over plotted as a cyan solid line, and  as a dashed line in the region where it has been extrapolated beyond the completeness limit of the LG.

 The background color scheme indicates the regions of the \mstar-\mhalo\ space where we expect to find cusps ($\gamma$$=$$-1.0$, in red) or progressively shallower cores (with cored-most galaxies having $\gamma$$=$$0.0$, in black).  
 Moving along the dark area in our map, we are moving in a region of constant \mstar/\mhalo$\sim$$0.004$, which is where we expect to find the cored most galaxies in our model. 
The bottom-right side of the plot is red, corresponding to the region of space with \mstar/\mhalo$\lesssim0.0001$: galaxies in this region do not have enough energy from stellar feedback to modify their profile, and they retain the initial NFW value.

Following the abundance matching line downwards from the top right corner, we are moving from a cored region, for galaxies of about \mstar$\sim10^8$\msun, to a cuspy area, for galaxies with \mstar$\lsim$$10^6$\msun, with galaxies in the middle populating the space of inner slopes $-0.2$$\lsim$$\gamma$$\lsim$$-0.8$.

In our model, Fornax and Sculptor are housed in haloes whose mass is \mhalo$\gtrsim10^{10}$\msun, and have enough stars to be within the core transformation region. In line with our findings, several  authors have argued that Sculptor and Fornax have a cored dark matter distribution (\citealt{Walker11,amorisco12,amorisco14b}, although \citealt{strigari14} have highlighted that a cuspy profile cannot be ruled out in Sculptor).

Fornax, with its inner slope of $\gamma$$=$$-0.39$, is the cored-most Milky Way satellite in our model that follows abundance matching predictions: it is probably the easiest satellite galaxy to reconcile with an energetic outflow scenario, given its high content in stellar mass \citep{penarrubia12,garrison-kimmel13}.
The Sculptor galaxy has a total halo mass of \mhalo$\sim1.5\times10^{10}$\msun\ and an inner slope of $\gamma$$\sim$$-0.65$: with such values its mass within 1.8\,kpc is about $\rm M_{1.8}$$\sim$$3.9$$\times$$10^8$\msun, in agreement with results from \citet{battaglia08}, who show that that observed velocity dispersion profiles are best fitted by a cored dark matter halo, and further show that an NFW profile yields a worse fit for the metal rich star population of Sculptor.

Leo I, being scattered to the right of the abundance matching line, lives within a relatively massive halo and falls in the mild-cusp region, $\gamma$$=$$-0.76$, in agreement with results from \citet{Mateo08}. The other Milky Way dSphs's that lie on the abundance matching line, Draco, Ursa Minor and Leo II, are housed in slightly smaller (\mhalo$\sim$$10^{9.8-10}$\msun), cuspy haloes. 
To summarize, we find that galaxies following the abundance matching of \citet{brook14} (and its linear extrapolation) do not form cores in haloes below \mhalo$\lesssim10^{10}$\msun.

As discussed in Section~\ref{environment}, our model for cusp vs. core is based on the analysis of simulated isolated galaxies, and does not take into account environmental effects such as tidal disruption and ram pressure stripping. We have thus excluded the satellites circled in red in Figure~\ref{fig:color_panel}, which may have been affected by their environment. In absence of environmental effects, the population of satellites living in the region of \mhalo$\lsim$3$\times$$10^9$\msun\  would likely be shifted to higher halo masses, with the cusp/core prediction changing accordingly. Whether such disrupted galaxies should have a core or not, cannot currently be determined by our model.

A more robust analysis can be performed for the isolated galaxies in our sample, as their kinematics are not affected by any environmental process.
We found that the expected cored-most galaxies should be WLM and IC1613. It has been argued that the hot stellar component of WLM may point toward a bursty star formation history that leads to a cored distribution \citep{teyssier13}: the  agreement between the distribution of stars and gas in controlled simulations  \citep{teyssier13,kawata14} and observations \citep{Leaman12} supports this scenario.

VV 124, Pegasus, Leo A and Aquarius live in dark matter haloes with inner slope $\gamma$$\sim$$-0.5$. Cetus and Tucana are relative outliers, scattered either side of the  abundance matching prediction. Cetus' kinematic implies that it lives in a halo with mass \mhalo$<$$10^{10}$\mstar, with inner slope $\gamma$$\sim$$-0.3$; on the contrary Tucana, with \mstar$\sim$$10^6$\msun, has kinematics the imply a significantly more massive halo, and therefore its inferred profile is cuspy, NFW.

\section{Conclusions}\label{conclusion}
We have fit an ensemble of 40 Local Group galaxies to dark matter haloes, based on their velocity dispersions and half-light radii, which provide an estimate of their dynamical mass enclosed within $r_{1/2}$ \citep{kirby14,wolf10}.
The haloes were selected from a LG  halo mass function which is well described by a single power law \citep{brook14}. In our fiducial normalization, the LG  halo mass function is set to have a maximum halo mass of \mhalo$=1.4\times10^{12}$\msun, in agreement with the most favored current mass estimates of the Milky Way and Andromeda galaxy, and resulting in a LG abundance matching that smoothly joins the large scale surveys \mstar-\mhalo\ relation \citep{guo10,moster13} in the region where the two volumes overlap.

Two different density profiles for dark matter haloes have been used: a cuspy NFW model \citep{navarro97} and a mass dependent DC14 profile \citep{DiCintio2014b}, which, taking into account the impact of baryonic processes on dark matter haloes, describes the formation of shallower dark matter distributions in galaxies whose stellar-to-halo mass ratio is 0.0001$\lsim$\mstar/\mhalo$\lsim$0.03 \citep{DiCintio2014a}. 
The resulting best fit halo mass is shown as a function of the  observed stellar mass for each galaxy in our sample, and compared with abundance matching predictions which have been empirically extended down to  \mstar$\sim$3-5$\times$10$^{6}$\msun\ \citep{brook14,elvis}, in a mass range where reionization does not leave any haloes dark.

Assuming an NFW density profile, the kinematics of the sample of galaxies are best fit by relatively low mass haloes, \mhalo$\lesssim10^{10}$\msun: this results  in a systematic disagreement with  LG  abundance matching predictions \citep{brook14,elvis} as well as with adjusted relations that take into account the reduction of dark matter mass by up to a $30\%$ caused by baryonic effects and the fact that some low mass halo may remain dark \citep{sawala14}.
A corollary of this mismatch is the existence of several haloes that are ``too big to fail", yet have not been assigned to any observed galaxy.

A LG halo mass function with maximum halo mass of \mhalo$=0.8\times10^{12}$\msun\ would reduce the number of ``too big to fail" haloes but, although within the limits of MW and M31 mass estimates, it would require that the most massive LG galaxies are systematically matched to low halo masses, compared to the galaxies observed in larger volumes \citep{guo10,moster13} such as in the SDSS and GAMA surveys \citep{baldry08,baldry12}; finally, even considering such a low normalization, the disagreement between kinematically-inferred NFW halo masses and  LG abundance matching predictions would remain. 
Whether such lower power law normalisation is the lowest possible, given the constraints  of the masses of M31 and MW, is beyond the scope of this paper, and to address this point a statistical sample of Local Group simulations is required. Therefore, given the cosmic variance, our work does not rule out the possibility of having only a particularly small number of haloes with masses that are ``too big to fail", although this issue was studied in detail in \cite{elvis}.

These results confirm previous works in showing that the steep and universal NFW density profile, when combined with the halo mass function expected from collisionless cosmological simulations, do not describe well the kinematics of LG  galaxies \citep{ferrero12,dicintio13,tollerud14,gk14,Ogiya14b,collins14}.

Regardless of LG mass, the NFW model places galaxies with $10^6$$\lsim$\mstar/\msun$\lsim$$10^8$ within haloes of $10^8$$\lsim$\mhalo/\msun$\lsim$5$\times$10$^9$, implying a reversal of the steep trend of decreasing star formation efficiency, \mstar/\mhalo, with decreasing halo mass that is found at larger stellar masses \citep{guo10,moster13}.
Considering the  effects of reionization and feedback on low mass galaxies, a reversal of such trend is difficult to explain within our current theoretical framework. Indeed, models of galaxy formation, both high resolution simulations \citep{brook12,munshi13,Shen14,hopkins14} and semi-analytic models \citep{Benson02,guo11}, predict  that star formation efficiency will continue to decrease steeply as halo mass decreases.

To explain the high star formation efficiency of LG  dwarf galaxies that arises from the NFW model, one would require a physical mechanism that allows such low mass haloes, \mhalo$\sim10^{8-9}$\msun, to have the same efficiency, in terms of \mstar/\mhalo, as haloes with mass \mhalo$\sim$$10^{11}$\msun. 
This reasoning is specially important for \textit{isolated} LG  galaxies, in which environmental effects can not be invoked to reduce their halo masses.

Yet one must consider the possibility of scatter in the \mstar-\mhalo\ relation. 
Observations of luminosity functions are incomplete below a certain minimum luminosity. 
This means that for any given halo mass we may be seeing the high stellar mass tail of the stellar mass distribution. Thus, although the average stellar-to-halo mass relation does continue to decrease, the only galaxies that we are able to observe at the low stellar mass end may not be average.  How much scatter exists in the \mstar-\mhalo\ relation, in particular for low mass galaxies, is an open question.

Remarkably different results are found when the kinematics of the sample galaxies are described by means of the mass dependent DC14 profile.

The most massive dwarfs, those with \mstar$\gsim$$3$$-$$5$$\times$$10^6$\,\msun, generally fit in haloes with mass \mhalo$\gsim$$10^{10}$\msun\, whose density profiles are shallower than NFW: the resulting distribution of halo masses is in agreement with abundance matching predictions and notably reduces the ``too big to fail" problem in the LG. 

The relation between \mstar\ and \mhalo\ for {\it isolated} dwarfs matches well the abundance matching relations of \citet{brook14,gk14} and their extrapolation below \mstar$\sim$3$\times$10$^{6}$\msun, which implies that LG  galaxies follow the same trend of decreasing star formation efficiency with decreasing halo mass as found for more massive galaxies in larger volumes \citep{guo10,moster10}.

Further, isolated galaxies in the DC14 model span a range of density profiles, moving from a cored region for the highest stellar mass objects, such as IC1613 and WLM, to a cuspy one for the galaxies with lowest stellar masses, such as LeoT and Tucana.
The  weak trend of $\rm V_{\rm max}$ versus \mstar\ pointed out in \citet{gk14} for the isolated galaxies in the LG  is explained in the light of this result:  a correlation between stellar mass and halo mass at scales smaller than \mstar$\sim10^8$\msun\  holds, but the density profiles of the haloes hosting such galaxies varies, with the brightest objects inhabiting the most massive, cored haloes, and the fainter galaxies living into smaller, cuspy haloes. 
Our findings therefore discard the existence of a universal profile within
the haloes that host LG galaxies, in agreement with previous studies \citep{collins14}.

About half of the {\it satellite} galaxies also follow the same  \mstar-\mhalo\ relation as the isolated galaxies, with the cored most Milky Way satellites expected to be Fornax and Sculptor, in agreement with previous studies \citep{Walker11,amorisco12,amorisco14b}.
The other half are instead fit to lower halo masses than predicted by abundance matching, likely as a result of environmental effects \citep{collins14}.

The fact that no isolated galaxies prefer to live in haloes with \mhalo$\lsim$5$\times$$10^{9}$\msun\ favors the scenario in which the environment have affected the kinematics of this subset of satellite galaxies: it seems likely that they where housed in more massive haloes prior to infall, and have had their masses reduced by the  effects of ram pressure stripping, baryon outflows and/or tides \citep{penarrubia10,zolotov12,arraki13,brooks13}. 

Asides from the signatures of tides that these galaxies show \citep{okamoto12,battaglia11,battaglia12}, another reason to favor environmental effects is that, if the mass of these objects was higher than their kinematics imply, their star formation efficiency \mstar/\mhalo\ can be reconciled with the one found amongst the isolated members of the LG.
Finally, acknowledging that some of the lowest mass (\mstar$\lsim3\times10^6$\msun) satellite galaxies may have been more massive in the past, helps solving the ``too big to fail" problem as well, as they would populate some of the haloes more massive than \mhalo$\sim7\times10^9$\msun. Future surveys, both blind HI as well as deep optical studies, will possibly identify some yet-undiscovered faint galaxy in the nearby universe further increasing the agreement with theoretical expectations.\\

The existence of a cusp/core transformation amongst LG galaxies is able to explain and reconcile \textit{at the same time} the measured shallow dark matter profile in some of the LG members, the ``too big to fail" problem, the abundance matching predictions and the star formation efficiency of such galaxies.

We have shown that a combination of feedback-driven halo expansion and environmental processes within a \LCDM\ context can account for the observed kinematics of Local Group galaxies, providing the theoretical framework to understand the inferred non-universality of their density profiles.

Detailed kinematical data of a larger sample, particularly of isolated galaxies, will provide crucial tests of our model.

\section*{Acknowledgements}
The authors thank the MICINN (Spain) for the financial support through the MINECO grant AYA2012-31101. They further thank Alyson Brooks and an anonymous referee for their thoughtful comments which improved the quality of the paper.
\bibliographystyle{mn2e}
\bibliography{archive}

\bsp


\label{lastpage}

\end{document}